\documentclass[twocolumn]{revtex4}
\usepackage{graphicx}
\usepackage{subfigure}
\begin{document}
\title{
	Experimental study for Yuen-Kim protocol of quantum key 
	distribution with unconditional secure
}
\author{
	Osamu Hirota, Kentaro Kato, and Masaki Sohma
	$^\dagger$\\
	Research Center for Quantum Communication,\\
	Tamagawa University, Tokyo, 194-8610, Japan.\\
	$\dagger$Matsushita Electric Industrial Co. Ltd, Japan\\
	E-mail:$\quad$hirota@lab.tamagawa.ac.jp
}
\begin{abstract}
	In this report, we simulate practical 
	feature of Yuen-Kim protocol
for quantum key distribution with unconditional secure. 
In order to demonstrate them experimentally
by intensity modulation/direct detection(IMDD) optical fiber 
communication system, we use simplified
encoding scheme to guarantee security for key 
information(1 or 0).
That is, pairwise M-ary intensity modulation scheme is employed. 
Furthermore, we give an experimental implementation of 
YK protocol based on IMDD.
\end{abstract}
\maketitle

\section{Introduction}
It is well known that quantum key distribution is one of the most
interesting subjects
in quantum information science, which was pioneered by C.Bennett and
G.Brassard in 1984[1]. In the original paper of Bennett, 
single photon communication was employed as implementation
of quantum key distribution. However, despite that it is not essential 
in great idea of Bennett, many researchers employed
single photon communication scheme to realize BB84, B92[2]. 
Because of the difficulties of 
single photon communication in practical sense, it was discussed 
whether one can realize a secure key distribution guaranteed by 
quantum nature based on light wave communication or not.

In 1998, H.P.Yuen and A.Kim[3] proposed another scheme for key distribution
based on communication theory(signal detection theory). 
This scheme corresponds to an implementation of secret key sharing
which was information theoretically predicted by Maurer[4], et al.
However, Yuen's idea was found independently from Maurer's discussion.
In the first paper of Yuen-Kim[3], they showed that if noises of
Eve(eavesdropper) and Bob(receiver) are statistically independent, 
secure key distribution can be realized
even if they are classical noises, in which they employed a modification of
B92 protocol[2].
Following YK's first paper, a simple experimental demonstration of 
YK protocol based on classical noise was reported[5], and recently 
YK scheme with 1 Gbps and 10 km long fiber system
based on quantum shot noise was demonstrated[6].
However, these schemes are not unconditional secure. That is, ability of
signal detection of Eve can be superior to that of Bob. 
As a result, an interesting question arises
" Is it possible to create a system with current technology that could
provide a communication in which
always Bob's error probability is superior to that of Eve?"

In proceedings paper of QCM and C 2002, Yuen and his coworker reported
that YK protocol can be unconditional secure, even if one uses conventional
optical communication system[7].
This is interesting result for engineer, and will open a new trend of 
quantum cryptography.

In this report, we simulate practical feature of Yuen-Kim protocol for
quantum key distribution with unconditional secure, and propose a scheme to
implement them using our former experimental setup[6].

\section{Yuen-Kim protocol}
\subsection{Basis}
A fundamental concept of Yuen-Kim protocol follows the next remark.\\
{\bf Remark}: {\it If there are statistically independent noises between Eve
and Bob, there exist a secure key distribution based on communication.}\\
They emphasized that the essential point of security of the key distribution
is detectability of signals.
This is quite different with the principle of BB-84, et al which are
followed by no cloning theorem.
That is, BB-84 and others employ a principle of disturbance of quantum
states to give a guarantee of security,
but YK protocol employs a principle of communication theory.
It was clarified that this scheme can be realized as a modification of B-92.
However, this scheme allows us use of classical noise, and it cannot provide
unconditional secure.
Then, Yuen and his coworker showed that YK scheme is to be unconditional
secure in which a fundamental theorem in quantum detection theory was used for his
proof of security as follows.\\
{\bf Theorem}: (Helstrom-Holevo-Yuen)\\
{\it Signals with non commuting density operators cannot be distinguished
without error.}\\
So if we assign non commuting density operators for bit signals 1 and 0,
then one cannot distinguish without error.
When the error is 1/2 based on quantum noise, there is no way to distinguish them. 
So we would like
to make such a situation on
process between Alice and Eve. To do so, a new version of YK scheme was
given as follows:
{\em
\begin{itemize}
\item[\rm(a)] The sender(Alice) uses an explicit key(a short key:$K$,
expanded into a long key:$K^*$
by use of a stream cipher) to modulate the parameters of a multimode
coherent state.
\item[\rm(b)] State
$|\Psi_0\rangle=|\alpha/\sqrt{2}\rangle_1\otimes|\alpha/\sqrt{2} \rangle_2$
is prepared. Bit encoding can be represented as follows:
\begin{eqnarray}
|\Psi_b\rangle
&=&
\exp\{-iJ_z\phi_b\}|\Psi_0\rangle
\nonumber\\
&=&
|e^{-i\phi_b/2}\alpha/\sqrt{2}\rangle_1
\otimes|e^{i\phi_b/2}\alpha/\sqrt{2} \rangle_2
\end{eqnarray}
where $J_z =({a^\dagger}_1{a_1} -{a^\dagger}_2{a_2})/2$.
\item[\rm(c)] Alice uses the running key $K^*$ to specify a basis from a set
of M uniformly
distributed two-mode coherent state.
\item[\rm(d)] The message $X$ is encoded as $Y_{K^*}(X)$. This mapping of
the stream of bits is the key to be
shared by Alice and Bob. Because of his knowledge $K^*$, Bob can demodulate from
$Y_{K^*}(X)$ to $X$.
\end{itemize}
}
\subsection{Security}
Here, let us introduce the original discussion on the security.
The ciphering angle $\phi_\nu$ could  have $k$ in general as discrete or continuous 
variable determined by distribution of keys. A ciphered two mode state may be
\begin{equation}
|\Psi_{bk}\rangle=\exp\{-iJ_z(\phi_b+\phi_k)\}|\Psi_0\rangle
\end{equation}
The corresponding density operator for all possible choices of $k$ is
$\rho_b$, where $b=1$ or $0$.
The problem is to find the minimum error probability that Eve can achieve in
bit determination. To find the optimum detection process for 
discrimination between $\rho_1$ and $\rho_0$ is the problem of 
quantum detection theory. The solution is given by[8]
\begin{equation}
<P_e> = \min_{\Pi}(p_1Tr \rho_1\Pi_0 + p_0Tr \rho_0\Pi_1)
\end{equation}
As an example of encoding to create $<P_e> = P_e(E)=1/2$ 
which is the error probability of Eve, Yuen et al suggested certain modulation scheme. 
In that case, closest values of a given $k$ can be
associated with distinct bits from the bit at position $k$,
and two closest neighboring states represent distinct bits which means a set
of base state. In this scheme, they assumed that one chooses 
a set of basis state(keying state for 1 and 0) for bits
without overlap. The error probability for density operators $\rho_1$
and $\rho_0$ becomes 1/2,
when number of a set of basis state increases.
Asymptotic property of the error probability depends on the amplitude of
coherent state[7][9].

\section{Simplified modulation scheme in YK protocol}
\subsection{Basic protocol}
Original scheme of YK protocol in the above can be realized by  practical
devices. To apply them to fiber communication system, 
we would like to realize them by intensity modulation/direct detection scheme.
If one does not want to get perfect YK scheme, one can more simplify 
the implementation of YK protocol.

From a fundamental principle in quantum detection theory, we can construct
 non-commuting density operators from sets based on non-orthogonal states 
when one does not allow overlap of the selection of a set of basis state for 1 and 0.
On the other hand, when we allow overlap for selection of a set of basis state,
one can use orthogonal state to construct the same density operators for 1 and 0.
That is, $\rho_1 =\rho_0$. However, in this case, unknown factor for Eve 
is only an initial short key, and a stream of bits that Eve observed is 
perfectly the same as those of Alice and Bob, though Eve cannot estimate 
the bits at that time. This gives still insecure situation. So, here, we employ
a combination of non-orthogonality and overlap selection in order to reduce 
the number of basis sets.

Let us assume that the maximum amplitude is fixed as $\alpha_{max}$.
We divide it into 2M. So we have M sets of basis state$\{(A_1,A_2),(B_1,B_2), \dots \}$. 
Total set of basis state is given
as shown in Fig.1. Each set of basis state is
used for $\{1, 0\}$, and $\{0, 1\}$, depending on initial keys.
\begin{eqnarray}
Set\quad A_1 : 0 &\rightarrow& |\alpha_{(1)}\rangle, 
\quad 1 \rightarrow |\alpha_{(M/2+1)}\rangle\\
Set\quad A_2 : 0 &\rightarrow& |\alpha_{(M/2+1)}\rangle, 
\quad 1 \rightarrow|\alpha_{(1)}\rangle
\end{eqnarray}
So the density operators for 1 and 0 for Eve are 
\begin{equation}
\rho_1=\rho_0 = \frac{1}{2}(|\alpha_{(1)}\rangle\langle \alpha_{(1)}|
+|\alpha_{(M/2+1)}\rangle\langle \alpha_{(M/2+1)}|)
\end{equation}
For the sets of $\{B_1,B_2\}$, $\{C_1,C_2\}$,$\dots $, 
let us assign 0 and 1 by the same way as Eqs(4),(5).
In this case, Eve cannot get key information, but she can try to know 
the information of quantum states used for bit transmission. 
So this is the problem for discrimination of 2M pure states.
The error probability is given by
\begin{equation}
<P_e> = \min_{\Pi} (1 - \sum p_iTr \rho_i\Pi_i)
\end{equation}
Although we have many results for calculation of optimum detection 
problems[10][11][12], to solve this problem is still difficult at present time, 
because the set of states does not have complete symmetric structure. 
So we here give the lower bound and tight upper bound.
The lower bound is given by the minimum error probability:$P^*_e(2)$ for signal set 
$\{|\alpha_{(1)}\rangle,|\alpha_{(2)}\rangle \}$ which are neighboring states. It is given as follows:
\begin{equation}
	P^*_e(2) = \frac{1}{2}\left(
	1-\sqrt{1-\exp[-|\alpha_1-\alpha_2|^2]}
	\right)
\end{equation}
The upper bound is given by applying square root measurement for 2M pure states.
The numerical properties are shown in Fig.2-(a). 
Thus if M increases, then her error for information on quantum states increases. 
In this case, pure guessing corresponds to $({2M-1})/(2{M})$. 
The error probability of Bob, however, is independent of the number of 
set of basis state, and it is given as follows:
\begin{equation}
P_e(B)= \frac{1}{2}(1 - \sqrt{1-|\langle \alpha_1|\alpha_{M/2+1}\rangle|^2})
\end{equation}
We emphasize that Eve cannot get key information in this stage, 
because the information for 1 and 0 are modulated by the way of Eqs(4),(5).
Furthermore, this scheme can send 2M bits by M sets of basis state.

Let us apply the original scheme such that M bits are sent by M sets of basis state.
In this case, Eve will try to get key information, so the density operators for Eve 
become mixed states $\rho_1$, $\rho_0$ consisting of set of states which send 1 and 0,
respectively.
The numerical properties are shown in Fig.2-(b). 
Both schemes have almost same security,
but the latter can only send M bits by M sets of basis state. 
In other word, the number of sets is reduced to 1/2 in the former scheme. 

\section{Primary design of experimental set}
In implementing YK protocol by conventional fiber communication system, 
we use here our proposed system. Figure 3 shows the experimental setup. 
The laser diode serves as 1.3$\mu$m light source. 
A pattern generator provides a signal pulse string to send keys.
A modulator which selects basis state follows 
a driver of laser diode.
The selector gives selection of amplitude and assignment of 1 and 0, 
and is controlled by initial keys.
The laser driver is driven by output signals of modulator. 
The optical divider corresponds to Eve. The case 1 is a type of 
``opaque", 
and the case 2 is a type of ``translucent". 
The channel consists of 10 Km fiber and ATT.
We can change the distance equivalently from 10 Km to 200Km by ATT.

The speed of pulse generator to drive laser diode is 311Mbps, 622Mbps, and 1.2Gbps.
The detector of Bob is InGaAs pin photo loaded by 50$\Omega$ register, 
and it is connected to an error probability counter which can apply to 12 Gbps.
The dark current is 7$nA$ and the minimum received power 
of our system is about -30 dBm.

In this system, the problem for degree of security is only power 
advantage of Eve which 
will be set in near transmitter(Alice).
When the eavesdropping is opaque, the error probability of Bob increases drastically, 
and the error probability counter shows almost 1/2, which means 
that the error of Eve is also 1/2. 
In this case, problem of communication distance 
is not so important.
We can detect the existence of Eve 
in any distance of channel.  
When the eavesdropping is translucent, Eve has to take only few power($\eta << 1)$ 
from the main stream of bits sequence in order to avoid the power level disturbance. 
In this case, the error of Bob does not increase. 
As a result, Alice and Bob cannot detect the existence of Eve.
The secure communication distance depends on the error probabilities of Bob and Eve.
Let $\kappa$ be transparency of channel from Alice to Bob. 
The detectability for Bob in this experiment setup depends on 
the signal distance(amplitude difference between two states as basis state):
$\kappa (\alpha_{max}/2)$ for 
$\{|\kappa\alpha_{(1)}\rangle,|\kappa\alpha_{(M/2+1)}\rangle\}$, 
and that of Eve depends on the signal distance:
$\eta \times \alpha_{max}/(2M)$. Here we assume that $\kappa =\eta$, and 
the total loss is 20dB which corresponds to 100 Km. Since our receiver requires about 
-30 dBm, the transmitter is -10 dBm. 
When M increases, sufficiently the error of Eve increases.

\section{Conclusions}
We examined a simulation of YK protocol based on intensity modulation/direct 
detection fiber communication system, and showed 
a design of implementation of
secure system based on our 
experimental setup which was used to demonstrate the first version of 
implementation of YK protocol. We will soon report complete 
demonstration in experiment by the above system.

\begin{figure}[p]
	\includegraphics[scale = 1]{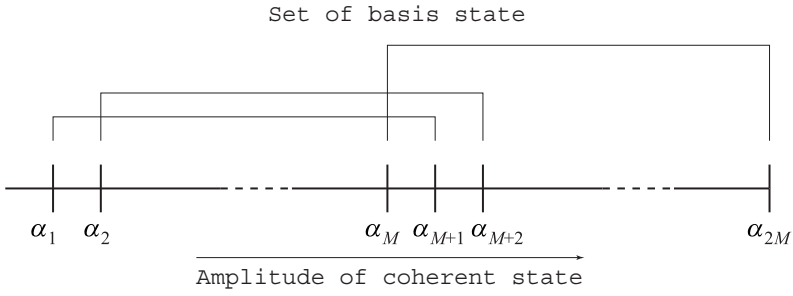}
\caption{
Selection of basis state
}
\label{fig1}
\end{figure}


\begin{figure}[ht]
  \begin{center}
      \subfigure[
      Error probability of Eve.
      Holizontal line: 
      upper is number of pure states $2M$;
      lower is number of set of basis state.
	]{%
        \includegraphics[scale = 0.9]{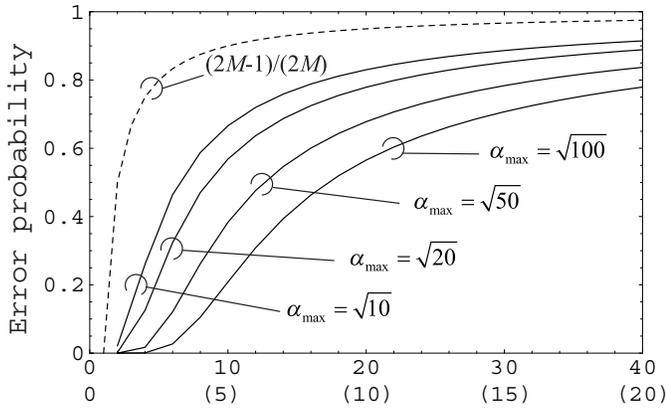}}\\
      \subfigure[
      Error probability of Eve.
      Holizontal line is number of set of basis state.
      ]{%
        \includegraphics[scale = 0.9]{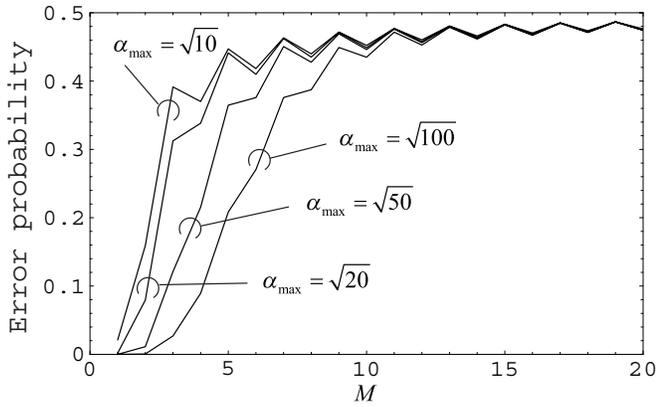}}\\
    \caption{Error probability of Eve}
    \label{fig2}
  \end{center}
\end{figure}

\begin{figure}[htb]
	\includegraphics[scale = 1]{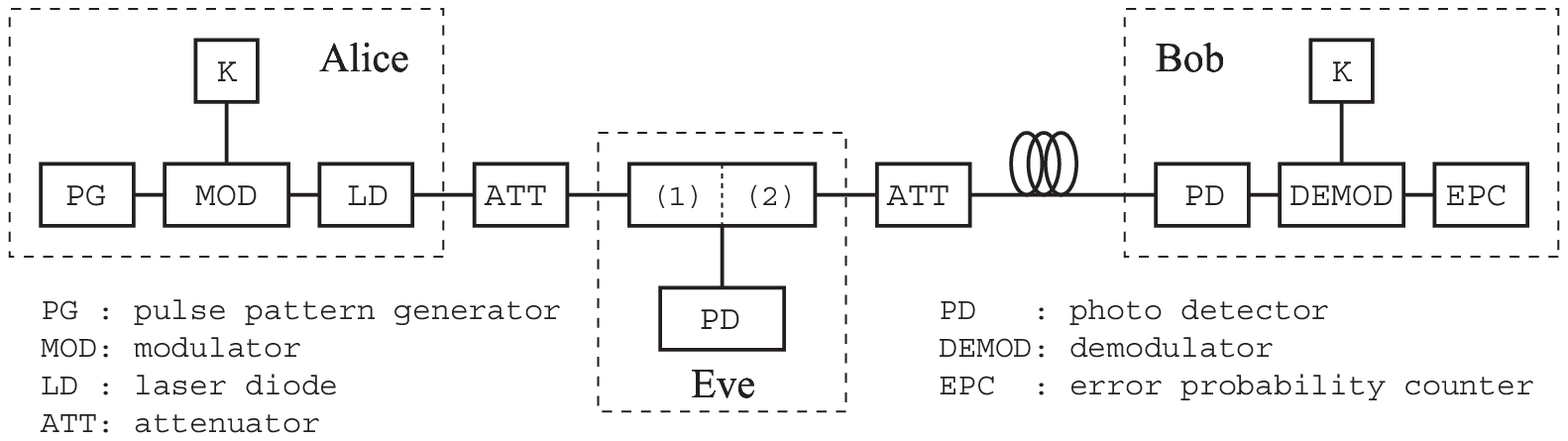}
\caption{
Experimental setup
}
\label{fig3}
\end{figure}

\begin{thebibliography}{99}
\bibitem{Bennett1} C.H.Bennett, and G.Brassard,
\newblock{in Proceedings of the IEEE International Conference on 
Computer, Systems, and Signal processing, India, 1984. pp175-179}.

\bibitem{Bennett2} C.H.Bennett,
\newblock{Phys. Rev. Lett., {\bf 68}, 3121, 1992}.

\bibitem{YK} H.P.Yuen and A.Kim,
\newblock{Phys. Lett. A, {\bf 241}, 135, 1998}.

\bibitem{Mau} U.M.Maurer,
\newblock{IEEE. Trans. {\bf IT-39}, 733, 1993}.

\bibitem{TH} A.Tomita, and O.Hirota,
\newblock{J. of Optics B: Quantum and semiclassical optics, {\bf 2}, 705, 2000}.

\bibitem{H} Y.Niwa, Y.Kudou, O.Hirota,
\newblock{unpublished, Report of Hitachi Shonnan Electron Co., 
and Tamagawa University, 2002}.

\bibitem{Ba1} G.A.Barbosa, E.Corndorf, P.Kumar, H.P.Yuen,
\newblock{e-print, LANL quant-ph/0210089, to be 
published in Proceedings of QCM-C, ed by J.Shapiro, Rinton Press, 2002}.

\bibitem{hel} C.W.Helstrom,
\newblock{Quantum detection and estimation theory, Academic Press, 1976}.

\bibitem{Ba2} G.A.Barbosa, E.Corndorf, P.Kumar, H.P.Yuen,
\newblock{e-print, LANL quant-ph/0212018,2002}.

\bibitem{Ban} M.Ban, K.Kurokawa, R.Momose, O.Hirota,
\newblock{International J. on Theoretical Physics, {\bf 36}, 1269, 1997}.

\bibitem{KOSH} K.Kato, M.Osaki, M.Sasaki, O.Hirota,
\newblock{IEEE, Trans. on Communications, {\bf COM-47}, 248, 1999}.

\bibitem{El} Y.C.Elder, and G.D.Forney,
\newblock{IEEE, Trans. Inform. Theory, {vol.47}, 858-872,
 Mar. 2001}.

\end{thebibliography}
\end{document}